%% file: ms.tex
\documentclass[12pt,preprint]{aastex}
\newcommand{\etal}{{et al.}}
\newcommand{\eg}{{e.g.,}}

\begin{document}

\title{Adaptive Optics Imaging Survey of Luminous Infrared Galaxies}

\author {Edward A. Laag\altaffilmark{1}, Gabriela Canalizo\altaffilmark{1}, 
Wil van Breugel\altaffilmark{2,3}, Elinor L. Gates\altaffilmark{4},
Wim de Vries\altaffilmark{2,5}, S. Adam 
Stanford\altaffilmark{2,5}}

\altaffiltext{1}{Institute of Geophysics and
Planetary Physics, University of California, Riverside, CA 92521}

\altaffiltext{2}{Institute of Geophysics and Planetary Physics, Lawrence 
Livermore National Laboratory, 7000 East Avenue, L413, Livermore, CA 94550}

\altaffiltext{3}{Department of Physics, University of California, Merced, CA 95344}

\altaffiltext{4}{Lick Observatory, P.O. Box 85, Mount Hamilton, CA 95140}

\altaffiltext{5}{Department of Physics, University of California Davis, 1 Shields Ave., Davis, CA 95616}

\begin{abstract}
We present high resolution imaging observations of a sample of previously 
unidentified far-infrared galaxies at $z<0.3$.  
The objects were selected by cross--correlating the $IRAS$ Faint Source 
Catalog with the VLA $FIRST$ catalog and the $HST$ Guide Star Catalog to allow
for adaptive optics observations.   We found two new ULIGs 
(with $L_{FIR} \geq 10^{12} L_{\sun}$) and 19 new LIGs (with $L_{FIR} 
\geq 10^{11} L_{\sun}$).
Twenty of the galaxies in the sample were imaged with either the Lick or Keck
adaptive optics systems in $H$ or $K'$.   Galaxy morphologies were
determined using the two dimensional fitting program GALFIT and the residuals 
examined to look for interesting structure.
The morphologies reveal that at least 30\% are involved in tidal
interactions, with 20\% being clear mergers.  An additional 50\% show signs
of possible interaction.
Line ratios were used to determine powering mechanism; 
of the 17 objects in the sample showing clear emission lines - four are active galactic nuclei and seven are starburst galaxies.  The rest exhibit a combination of both phenomena.
\end{abstract}


\section{Introduction}

Luminous infrared galaxies (LIGs; $L_{IR} \geq 10^{11} L_{\sun}$), 
are the best candidates for a link 
between the more powerful ultra-luminous infrared galaxies (ULIGs; $L_{IR} \geq 10^{12} L_{\sun}$) and 
normal quiescent elliptical galaxies \citep{gen01}.  
ULIGs are often interpreted as powerful mergers of gas rich spiral galaxies.
Good evidence exists for a correlation between IR luminosity and the
fraction of galaxies which are interacting \citep[\eg][]{san96}.
At $L_{IR} < 10^{11} L_\odot$ most IR galaxies are single,
gas--rich galaxies powered by normal star formation, while at $L_{IR}
> 10^{11} L_\odot$ there is a large increase in the fraction of
strongly interacting or merging galaxies and an increase in
the fraction of AGN--powered galaxies.  
Results of numerical simulations by \cite{mih94} describe the evolution of global star
formation rate (SFR) for merging pairs of disk + bulge + halo galaxies.  
At the first close approach, star formation
is shown to increase slightly over normal levels.  When the galaxies finally collide, gas is 
driven into the compact center of the 
remnant galaxy and the SFR increases rapidly to a peak as much as
70 times the initial rate.  

A population of ULIGs 
at high $z$ have been shown to share some of the properties of the local 
population by studies with {\em Infrared Space Observatory (ISO)} \citep[\eg][and references therein]{san02} and with the SCUBA camera on the James Clerk Maxwell Telescope (JCMT) \citep[\eg][and references therein]{bar99}.  If 
that is the case, then ULIGs probably played an important role in the star formation history of the universe.

Several high resolution imaging studies of ULIGs and LIGs have been conducted with the {\em Hubble Space Telescope (HST)}.
In particular, \citet{bor99} observed 120,
$z < 0.2$ ULIGs in snapshot mode using the F814W $I$--band filter and
found that virtually all of the objects in their sample are interacting or merging, and
that as many as 20\% contain multiple nuclei or are dense groupings of
interacting (soon--to--merge) galaxies. \cite{far01} observed a sample of 23 ULIGs with the $HST$ WFPC2 camera in $V$ band, and found 87\% to be interacting.  Borne \etal\ argue, as have
others, that there may be an evolutionary progression from compact
galaxy groups to galaxy pairs to ULIGs to elliptical galaxies. These
$HST$ data also reveal unresolved nuclei, probably AGN, in 15\% of the
objects, in good agreement with optical and FIR spectroscopic
classifications.  

More recently ULIGs and LIGs have been studied with ground based telescopes.  \cite{vei02} observed a sample of 118 ULIGs with the University of Hawaii 2.2m telescope in $R$ and $K'$.  Optical spectroscopy for this sample was published in \cite{vei99}.  They find virtually 100\% of the sample to be interacting, 39\% to be in the early stages of merging and  56\% to harbor a single disturbed nucleus in the late stages of a merger.  5\% were found to be multiple mergers.  They find 35\% of their surface brightness profiles to be fit well by a pure de Vaucouleurs $R^{1/4}$ profile and another 38\% to be fit equally well by either an exponential or de Vaucouleurs profile.  Mean half-light radius for their ULIGs was found to be $\sim$3.5 kpc in $K'$.

Considerable observational effort has been directed at
determining whether ULIGs and LIGs
are powered mainly by starbursts
or active galactic nuclei (AGN), what the evidence is for morphological 
evolution, and whether
these correlate with far-infrared (FIR) luminosity or spectroscopic classification
(starburst, AGN, LINER).
Recent spectroscopic 
surveys have shown that most of the FIR galaxies 
seem to be powered by
starbursts, but that the fraction of AGN-powered galaxies increases
with FIR luminosity \citep[\eg][]{vei99}.    
A morphological merger sequence which correlates with these spectroscopic
classifications is not clear, most likely because of large differences
in the time scales for the various events.  
\cite{arr04} observed 30 LIGs with the Nordic Optical Telescope in the visible bands $B$, $V$ and $I$.  They find that the LIG population is dominated by starbursts while a higher proportion of ULIGs are dominated by AGN activity and could actually evolve into QSOs.



In an effort to construct larger samples of LIGs and ULIGs having high
resolution imaging in the near IR, we have identified a new set of candidate
objects following the method of \citet[][]{sta00} as described below.  In 
addition, we have cross-correlated this sample with a bright star catalog so
as to allow for adaptive optics (AO) observations of the sample.   The main 
aim of our study was twofold: 1) to identify LIGs and ULIGs at higher 
($z > 0.1$) redshifts than current FIR-selected samples, and 2) to perform a
detailed high resolution morphological study that would allow us to identify
morphological sequences and characterize galaxy interactions in these objects.

\section{Sample Selection}

We have constructed a sample 
 by cross-correlating the $FIRST$ catalog \citep[$S_{\rm 1.4
GHz} > 1$ mJy, $5\sigma$ with 5\arcsec\ resolution;][]{bec95} 
with the $IRAS$ Faint Source Catalog 
\citep[$FSC$, $S_{\rm 60\mu m} > 0.2$ Jy, $5\sigma$;][]{mos92}. 
The sky coverage for the $FIRST$ catalog is roughly from RAs 8 to 17 hrs.

We extracted all sources which were optically faint, as detailed in
\citet{sta00}.  For a flux-flux plot and a plot of radio power at 1.4 GHz 
versus FIR luminosity 
illustrating the entire cross-correlated $FIRST$--$FSC$ sample
see figures 1 and 4 in \citet{sta00}.  
A major advantage of choosing these $FIRST$--$FSC$ (FF\,) matches is that it also 
provides a good reason to believe that the FIR flux comes from the optical 
object at the radio source position within the large $IRAS$ error ellipse. 

The FF\, sample was further cross-correlated with
 the $HST$ Guide Star Catalog (GSC) to define a 
sub--sample of LIG/ULIG candidates within one arcminute of stars of
magnitudes brighter than 13 in $R$.  The nearby stars can be used as guide stars
for observations with the Lick and Keck AO systems.   This yielded a sample of $\sim100$ targets.

Since the aim of this study was to identify new LIGs and ULIGs, we only considered those objects for
which no published redshift was available at the time (although the redshift of
roughly half of the objects in the sample has been published since by  
the Sloan Digital Sky Survey and other authors).  The cross-correlation of the three catalogs yielded a sample of $\sim50$ targets with RAs between 8 and 17 hrs that were previously unidentified.   
Due to observing constraints and weather conditions, we were only able to 
obtain redshifts for 28 objects.   Of these, two are found to be ULIGs, 19 are LIGs,
and the remaining seven we simply designate as IR galaxies (IRGs) with FIR 
luminosities $10^{10} \leq L_{FIR} < 10^{11} L_{\sun}$.  

Finally, we obtained Lick and Keck AO images of 20 of these objects as
described below.  The  sample is given in Table~\ref{sample}.  The 20 objects imaged at Lick 
and Keck observatories appear first in the table, in order of increasing 
RA.  The additional 8 galaxies, for which we have only spectra, are listed 
below the horizontal line. 

Column 1 is the target galaxy with its 
$FIRST$--$FSC$ catalog name.  Columns 2 and 3 list, respectively, the J2000.0 RA and DEC 
of the target.  Column 4 is the redshift obtained from the object spectrum.  
Column 5 is the $60\mu$m flux.  
Column 6 is the $100\mu$m flux; values in parentheses indicate upper limits.
The fluxes are the template amplitudes from the ``1002" median scans obtained with the SCANPI utility at IPAC.  Column 7 is the luminosity distance, calculated using $h=0.71$, 
$\Omega_{\Lambda}=0.73$ and $\Omega_{m}=0.27$ (which we assume throughout the
paper).  Column 8 is the integrated flux at 1.4 GHz.  Column 9 gives the FIR 
luminosity, calculated as in \citet{sta00}.  ULIGs and LIGs are normally defined according to $L_{IR}$ (8-$1000\mu$m) as a whole.  We have based our ULIG and LIG definitions on $L_{FIR}$ because in the majority of cases we only have $IRAS$ detections at $60\mu$m and $100\mu$m for the objects in our sample.  Most of the objects in the sample were not firmly detected by IRAS at $12\mu$m or $25\mu$m.  As a consequence it is possible that some of the objects we have 
classified as LIGs may actually be ULIGs.

The definition for $L_{FIR}$ used is from \cite{san96} and takes the form
\begin{equation}
L(40-500 \mu m) = 4\pi D_{L}^{2}CF_{FIR}[L_{\sun}] 
\end{equation}
where $D_{L}$ is the luminosity distance in Mpc, 
\begin{equation}
F_{FIR}= 1.26\times10^{-14}(2.58\times f_{60} + f_{100})[W m^{-2} ]
\end{equation}
and $C=1.6$.


\input{tab1.tex}

\section{Observations and Data Reduction}
Spectroscopic observations of each galaxy in the sample were obtained
using the Kast Double Spectrograph at the cassegrain focus of the Shane
3-meter telescope at Lick Observatory.   We used the 600/4310 grism for the
blue side and the 600/7500 grating for the red side to obtain a useful
wavelength coverage spanning from the atmospheric cutoff at $\sim$ 3400~\AA\ 
to 8100~\AA .
We used different slit widths to match the seeing conditions, typically
between 1\arcsec\ and 2\farcs5, yielding a resolution between 2.6 and 
6.5 \AA\ pixel$^{-1}$ for the blue side and between 3.3 and 8.3 \AA\  
pixel$^{-1}$ on the red side, so that the typical resolution for the spectra 
was roughly 300 km s$^{-1}$.   The total integration time for each galaxy was 
900~s.

The spectra were reduced with IRAF, using standard reduction procedures.  
After correcting for bias and flat fielding, we subtracted the sky and
wavelength-calibrated the two-dimensional spectrum using OH skylines (for 
the red side) and arc lamps (for the blue side).  We then flux-calibrated the 
spectra using
spectrophotometric standards from \citet{mas88} and extracted the 
spectra using the IRAF apextract routines.  In the majority of cases, we 
measured redshifts from stellar absorption lines, so that the redshifts we quote
correspond to the stellar component of the galaxies as opposed to the gas.

The galaxies FF\,1122+4315 and FF\,1429+3146 were imaged in $K'$
using the Keck II AO system \citep[][]{wiz00a,wiz00b,joh00}
with the NIRC-2 camera (PI: K. Matthews \& T. Soifer).
Both galaxies were observed using the NIRC-2 Wide-Field camera, which
yields a plate scale of $0\farcs04$ pixel$^{-1}$. In addition, 
FF\,1429+3146 was also observed with the Narrow-Field camera, which yields a 
plate scale of $0\farcs01$ pixel$^{-1}$.
The remaining galaxies in the sample were observed using the natural 
guide star LLNL AO system on the 3 meter Shane telescope at Lick Observatory;
for details about the LLNL AO system refer to \citet{bau99} and 
\citet{gav00}. The AO system feeds the AO-optimized infrared camera IRCAL 
\citep[][]{llo00}, yielding a plate scale of $0\farcs076$ pixel$^{-1}$. 
Observations at 
Lick were done in $H$ rather than $K_{s}$ since the warm optical elements in 
the AO system result in a high thermal background in the latter.
The AO FOV is 20\arcsec\ for the Lick system, 10\arcsec\ for the Keck Narrow-Field camera and 40\arcsec\ for the Keck Wide-Field camera .

Observations of point spread function (PSF) stars were obtained either immediately before or after the observations for each galaxy.  We attempt to account for anisoplanatism by matching the distance and position angle from the guide star (GS) to the PSF.  However, atmospheric conditions vary somewhat on a shorter scale than our total integration times, so that each PSF is close, but does not perfectly match the conditions for each image.  For this reason, we are unable to provide precise Strehl ratios for each image, but we estimate that the typical ratios for all our images were between 0.1 and 0.2.   We provide full width at half maximum (FWHM) of the PSF for each image as an indication of the system performance for every field.  Only the observations at Keck and those at
Lick on 25 Jan 2003 were done under photometric conditions; we estimate an 
extinction in $H$ between 0.1 and 0.4 magnitudes for the rest of the 
observations.  The images were reduced with IRAF, using standard IR reduction procedures.   A complete journal of observations is given in Table~\ref{journal} which includes total exposure times and GS information.

\input{tab2.tex}

\section{Analysis}
\subsection{Fitting Technique}

Fitting a mathematical model to an image of a galaxy is one 
consistent way of determining its morphology.  Several different mathematical 
models have been used over the years to fit the most common galactic shapes.  
These include the well known de Vaucouleurs profile that models elliptical galaxies and the exponential profile for galactic disks.  More recently the \cite{ser68} model has become a highly valuable tool for modeling various components of galaxies including bulges, disks, and bars.  Using the S\'{e}rsic model is beneficial because the model is able to adapt to the de Vaucouleurs (elliptical) profile at S\'{e}rsic index N=4, the exponential disk profile at N=1, and a Gaussian shape at N=0.5.

Many authors have used a one dimensional ellipse fitting routine to plot a light profile such as IRAF {\it ellipse}.  For example, \cite{vei02} use the standard Fourier expansion of \cite{bin98} to fit isophote ellipses to their galaxies.   After generating these ellipses the programs then make a plot of isophote intensity versus radius and derive the surface brightness profiles by making a best fit to those points.
Surveys based on these routines usually classify the object as either elliptical or disk shaped based on whether a de Vaucouleurs or exponential model fits best.  While this is not unreasonable, some of these classifications now need to be revised because a one dimensional fit can be subject to errors due to isophote twists and the large variety of galaxy morphologies.  Some of these surveys use a profile slice along the major or minor axis and some use both.  The profile can be different depending on whether the major or minor axis is used \citep[see][and references therein]{pen02}.  

The galaxies in this sample were fit in two dimensions using a program called GALFIT \citep{pen02}.   The program bypasses ellipse fitting, and fits a model to the light profile directly.   GALFIT uses $\chi^{2}$ fitting to minimize the error in two dimensions and can handle the fitting of multiple models simultaneously.   The best fit is then subtracted from the data, and the residuals can be analyzed.  In this survey, one or two models were used as needed to make a reasonable subtraction.  More models can be used, but improvements to the subtraction are not always desirable.   Generally, as resolution improves, many S\'{e}rsic profiles can be fit to a galaxy at the same time but these are not necessarily physically meaningful.  Depending on the number of components, a S\'{e}rsic model can have anywhere from eight to several dozen free parameters. 
 
The S\'{e}rsic parameter N affects the degree of cuspiness of the galaxy.  
As the S\'{e}rsic index decreases, the galaxy becomes more ``cuspy" in the center.  This means the core intensity flattens quickly as r increases to the half-light radius and the intensity falls off steeply beyond the half-light radius \citep{pen02}.  Unperturbed spiral galaxies are a homogeneous group.  Their S\'{e}rsic index usually does not deviate from N=1.  Similarly, ellipticals do not deviate much from N=4, though they are a less homogeneous group.   Mergers of two disk galaxies tend to produce a bulge with knots of compact unresolved star formation and some mass concentrated in the center.  In general, these types of mergers can have S\'{e}rsic indices that are greater than 4 or other ambiguous profiles.

\subsection{Fitting Procedure}

The fitting process consists of two stages.  The first stage involves an 
initial guess.  A visual evaluation of the galaxy is made to decide what types of objects are present (disks, bulges, bars, etc).  The general parameters of these objects, such as effective radius, centroid, ellipticity and position angle can all be approximated by examination of the image with an image analysis tool.  

The fitting program uses the initial guess parameters to create a model, and convolves it with a PSF provided by the user in order to match the seeing conditions and the resolution.  A ``best fit'' model is obtained by performing a least squares fit of the PSF-convolved model to the data.
The second stage consists of refinements to the model to achieve the most accurate representation of the galaxy.  The accuracy of the model can be assessed by examining the residual image for artifacts of under- or over-subtraction.  The residual image is the image formed when the model is subtracted from the original image.  If the galaxy is simple and unperturbed then the residuals should be only the sky background or spiral arms for spiral galaxies.  This is especially true for distant galaxies where the resolution that we achieve prevents us from resolving extreme detail such as globular clusters.

If the galaxy is perturbed then other types of residuals will be apparent.  These may consist of dust lanes, tidal tails, or multiple nuclei.  Large symmetric areas with negative values
are characteristic of over-subtraction.  These types of errors are generally easy to spot and another fit should be considered.   Sometimes the model may be obviously wrong.  Many adjustments to the input parameters can be made to refine the model.  One possibility is isolating variables by fixing quantities that are known to be accurate and letting the program fit a smaller subset of the parameters.   

The accuracy of the parameters determined by the fitting procedure will naturally depend on how well the PSF matches the conditions under which a given image was taken.    To go beyond a simple determination of disk versus elliptical and look at more complicated features, we need to evaluate the uncertainty in the model due to the PSF.   We ran several tests where we used a ``mismatched" PSF to fit various galaxies.   The mismatched PSFs were obtained with the same instrument and set up, but under different  observing conditions and with FWHM that differed by as much as 50\% of the value of the actual corresponding PSF.  We found that, overall, the resulting model parameters changed by only a small percentage of their value.    In particular, S\'{e}rsic indices of N $>$ 1 changed by less than 5\% and the position angle by only a few degrees.   Although smaller S\'{e}rsic indices (N $<$ 0.5) varied by a larger percentage (as much as 40\%), their values remained close to the values for disk profiles.  The effects are similar to running the program with no PSF.  It should be noted that in some cases the PSF can be so ``bad", for instance when position angle differs greatly from the image, that GALFIT will not be able to converge. In such cases it is better to run the program with no input PSF.   We therefore feel confident that, while the precise value of the S\'{e}rsic index may be uncertain due to uncertainties in the PSF, the overall determination of the galaxy morphology and global features is robust.  A more serious concern when a PSF is mismatched is that the structure in the residuals will be affected.   If the PSF does not properly represent the image quality of the galaxy, fine structure in the residuals will probably be lost.     

Another factor that will affect the S\'{e}rsic parameter is surface brightness fading.   For galaxies at higher $z$, the disk of a galaxy (where there is less signal) will fade faster than the central regions causing the galaxy to appear significantly smaller. 
This is because flux scales roughly as
$L/z^{4}$,
where $L$ is the luminosity measured at the source and $z$ is redshift 
\citep{mis73}.
In the case where there is a disk + bulge, losing the edges of the disk will increase the cuspiness and raise the S\'{e}rsic index.  

Finally a limited field of view may effectively raise the S\'{e}rsic index.  
When combined with surface brightness fading, a situation is created 
where a galaxy appears to be smaller and the drop off from the center is 
steeper.  In such a case only the central bulge of a galaxy is modeled.  For 
nearby galaxies where the field of view is less than the scale length 
$r_{s}$, the model for the galaxy may not be correct \citep{pen02}.   Even for our nearest galaxy, FF\,0841+3557,  at $ z = 0.036$ the FOV for the Lick AO system corresponds to 14 kpc.   For the exponential profile $r_{s} = (1.678r_{\frac{1}{2}})$, and if we take FF\,0841+3557 as an example, we calculate $r_{s}$ to be 7.7 kpc.

\begin{figure}[htb]
\epsscale{0.7}
\plotone{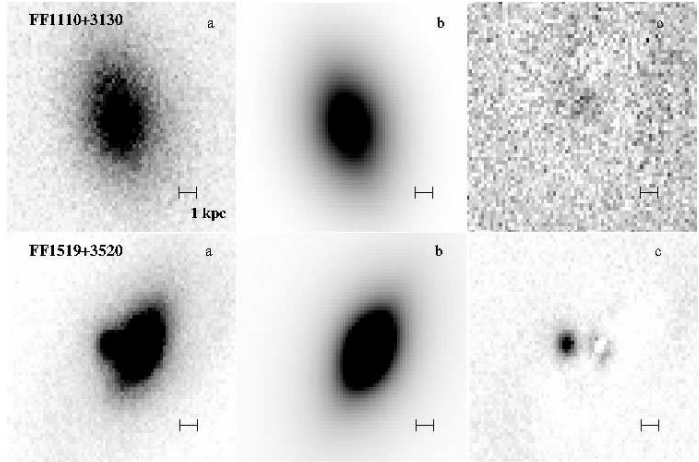}
\caption{For both rows, panel $a$ is an image of the galaxy before subtraction.  Panel $b$ is a numerical model of the galaxy and panel $c$ is the residual image produced from the subtraction of panel $b$ from $a$.  In this and the following figures, north is up, east is to the left, and the scale bar represents approximately 1 kpc.
\label{ff11101708}}
\end{figure}

Figure~\ref{ff11101708} is an example of how a residual image is obtained for two galaxies in the sample.  The top three panels, showing FF\,1110+3130, are an example of a numerical model that is an accurate representation of the morphology of the galaxy.  It produces a clean residual which shows sky background.  Panel $a$ is an image of the galaxy before subtraction.  Panel $b$ is a model of the galaxy and panel $c$ is the residual image produced from the subtraction of panel $b$ from $a$.  
The bottom three panels, showing FF\,1519+3520, are an example of a reasonable 
fit with interesting residuals, namely a compact companion $\sim0\farcs 5$ east of the galaxy nucleus.  In this case we have only modeled the larger 
galaxy to the west. 


\subsection{Results}

\begin{figure}[ht]
\epsscale{1.0}
\plotone{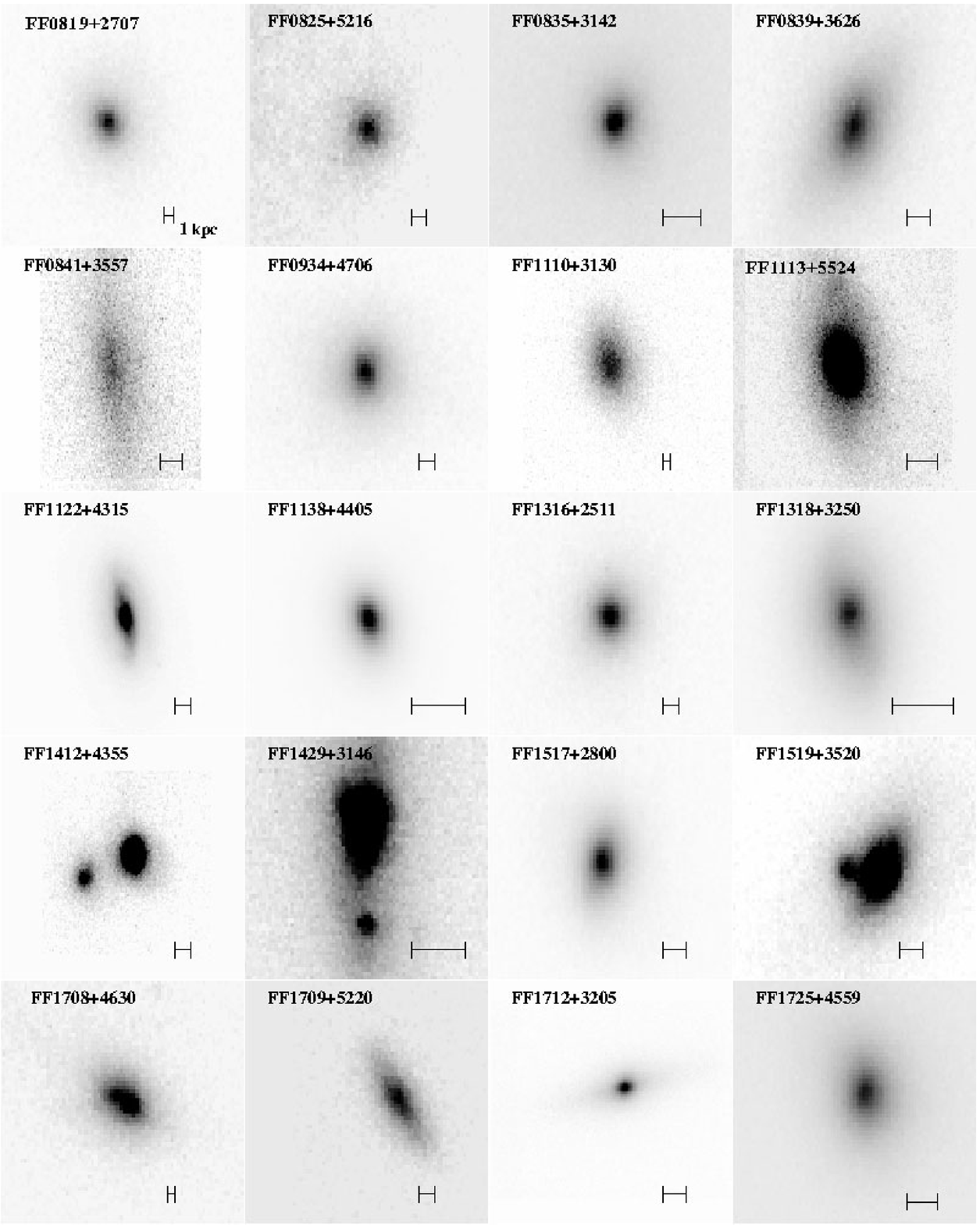}
\caption{Lick and Keck near IR AO images of the central regions of each galaxy in the sample.  
\label{msgals}}
\end{figure}

Figure~\ref{msgals} shows the $H$ or $K'$-band image of each of the 20 
galaxies imaged in the sample.  In terms of FIR luminosity, the 20 imaged 
objects consist of two ULIGs, 13 LIGs and five IRGs. The additional 
eight galaxies for which we only have spectra yielded six LIGs and 
one IRG.

For each of the galaxies modeled in the sample, the parameters for their models are listed in Table \ref{parameters}.
Column 1 lists the target galaxy which may have one or two component objects that are modeled within it.  Column 2 lists the apparent $H$ or $K'$ magnitude of the object being modeled as determined by the best fit.  Column 3 is the S\'{e}rsic parameter, N.  When the model used was an exponential profile the value of the S\'{e}rsic index is by default set to 1.  
The S\'{e}rsic model or ``light profile'' is given by:
\begin{equation}
\Sigma(r) = \Sigma_{e} e^{-\kappa[(r/r_{\frac{1}{2}})^{1/N} - 1]}
\end{equation}
where $ \Sigma(r)$ is the surface brightness at a given radius $r$, $r_{\frac{1}{2}}$ is the effective radius of the galaxy, $\Sigma_{e}$ is the surface brightness at $r_{\frac{1}{2}}$, N
is the S\'{e}rsic index and $\kappa$ is coupled to N so that half of the total flux is always within $r_{\frac{1}{2}}$.
This model is described in detail in \cite{pen02}. The paper should be consulted for the mechanics of how the model works, because more equations are involved than what is shown here.
The exponential profile is given by:
\begin{equation}
\Sigma(r) = \Sigma_{0} e^{-(r/r_{s})}
\end{equation}
where $r_{s}$ is the scale length $(1.678r_{\frac{1}{2}})$.
Column 4 lists the half-light radius ($r_{\frac{1}{2}}$) of the object in kpc.  
Column 5 is the ellipticity ($\varepsilon$) of the isophotes (defined as the $1 - b/a$ where b/a is the axis ratio for an ellipse).  Column 6 is the position angle in degrees.
Column 7 is the boxiness-diskiness parameter.  A negative value corresponds to a disky galaxy and a positive value corresponds to a boxy galaxy (see Section 4.4 below).
Column 8 is the model type used for the object where ``S\'{e}r" is short for S\'{e}rsic and ``Exp" is short for exponential.  Column 9 indicates whether the galaxy appears to contain a single nucleus, a double nucleus, or  multiple nuclei.  

Column 10 indicates whether the galaxy is interacting or not.  We classify a
galaxy as interacting (``Y'') if the galaxy 
appears to be in the early stages of a merger (i.e., still showing 
distinct components) and/or shows obvious tidal tails or other tidal debris.
We designate a galaxy as possibly interacting (``?'') if the galaxy was 
difficult to fit, implying that it may have an irregular morphology.
The rest we designate as non-interacting (``N'').  The models used for each 
individual object are described in detail in Section 5 below.

\subsection{Powering Mechanism}
 
ULIGs and LIGs can be classified in two categories according to the primary source of their luminosities: those galaxies that achieve their high luminosity from starburst activity (henceforth referred to as a starburst galaxies) and those that are mainly powered by AGN.    It is plausible that most LIGs contain some combination of excitation mechanisms including AGN, starbursts, shocks, mergers, and bars.  The question then becomes which mechanism is dominant in each galaxy and whether there are any trends evident in the sample.  
 
\begin{figure}[htb]
\epsscale{0.7}
\plotone{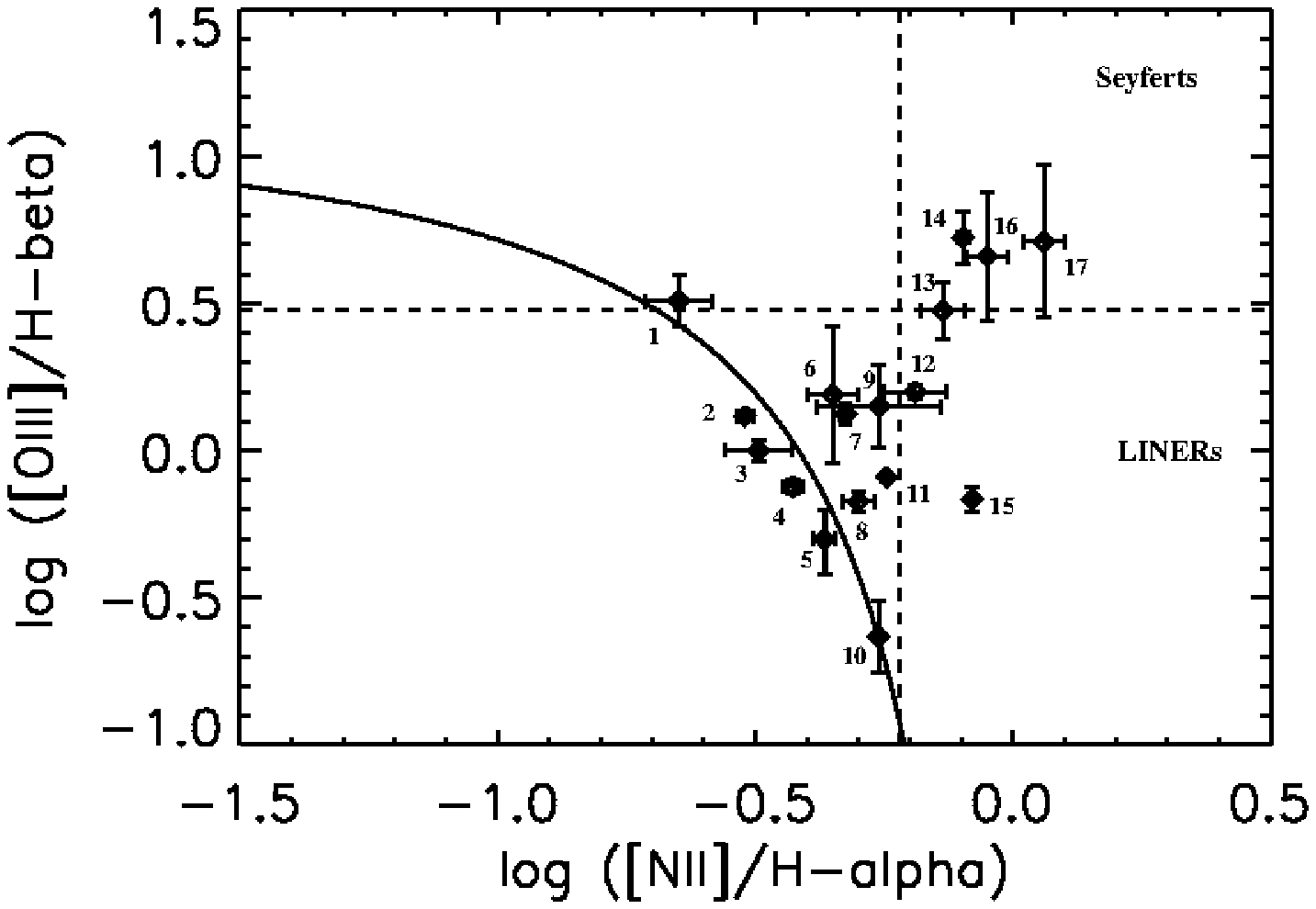}
\caption {A BPT diagram of emission line flux ratio [O\,III]/H$\beta$ versus the ratio [N\,II]/H$\alpha$. From left to right the
galaxies plotted are: (1)~FF\,1656+2644, (2)~FF\,1110+3130, (3)~FF\,1709+5220, (4)~FF\,0934+4706, (5)~FF\,1712+3205, (6)~FF\,1721+2951, (7)~FF\,1517+2800,
(8)~FF\,1138+4405, (9)~FF\,1723+3845, (10)~FF\,1318+3250, (11)~FF\,1412+4355, (12)~FF\,1429+3146, (13)~FF\,0834+4831, (14)~FF\,1519+3520, (15)~FF\,1651+3001, (16)~FF\,1122+4315, (17)~FF\,0835+3142.
\label{agnid}}
\end{figure}

Figure \ref{agnid} shows the emission line flux ratio [O\,III]/H$\beta$ versus the ratio [N\,II]/H$\alpha$ plotted for the 17 objects in the sample which
had firm detections of all four emission lines and a redshift $z\lesssim0.2$ (so that both H$\alpha$ and 
H$\beta$ are in our observed spectral range).  
This plot is similar to the plot in Fig.~1 of \cite{kau03}
which itself is derived from the ``BPT diagram". \citet{bal81} demonstrated
that it is possible to distinguish type-2 AGNs from normal star forming galaxies by plotting their emission line ratios and this idea was expanded upon by 
others.  The curved line represents the demarcation between starburst- and AGN-powered 
galaxies as determined by \citet{kew01}. 

According to the diagram, the galaxies FF\,0834+4831, FF\,0835+3142, FF\,1122+4315, and FF\,1519+3520 are identified as AGN, while  FF\,0934+4706, FF\,1110+3130, FF\,1656+2644, FF\,1709+5220,  FF\,1712+3205, and FF\,1318+3250 are identified as starburst galaxies.  Additionally, FF\,1138+4405 lies near the demarcation line, and is likely a star forming galaxy as well.   
In both of these groups, there is a range of FIR luminosities so that we do 
not see a clear correlation between $L_{FIR}$ and powering mechanism for the
LIGs and IRGs in the sample, although it is important to note that some of
the objects are classified as LIGs based on upper limits to $L_{FIR}$ and this
adds scatter to any possible trends.  Neither of the two ULIGs are plotted in 
Fig.~\ref{agnid} because [N\,II] and H$\alpha$ were redshifted out of our
observed spectra.   However, they both have [O\,III]/H$\beta$
ratios characteristic of AGN.  
Studies with larger samples like \cite{vei99} and \cite{arr04} indicate the fraction of AGN dominated LIGs increases with $L_{FIR}$.  FF\,1651+3001 is the only galaxy that falls in the LINER region. The rest of the objects fall in a region intermediate between starburst and AGN, possibly representing a population containing some combination of both AGN and starbursts, with neither being clearly dominant.

\input{tab3.tex}

\subsection{Morphologies}
 
One of the two ULIGs, FF\,1708+4630, is a merger at an early stage, while the 
other one,
FF\,0819+2707, has possible signs of interaction.
Three LIGs, FF\,1412+4355, FF\,1429+3146 and FF\,1519+3520, are also mergers
at an early stage, where the two nuclei are still distinct.
The LIG mergers are likely mergers of two disks since both components can be 
fit well by near exponential profiles and are in  
early merging stages, because both nuclei are still distinct.  
Between 30 and 60\% of LIGs are reported to be mergers in the literature 
\citep{san96}.  Similarly, 23\% (3/13) of the LIGs in our sample are found to be mergers. In agreement with the rarity of multiple mergers cited by \cite{vei02}, we find only one object FF1429+3146 (a LIG) that appears to be a multiple merger in the residual image.  None of the 5 IRGs have multiple nuclei or obvious signs of interaction.  

There are only 2 objects that seem to have large bulge components in the sample.  One of these objects is the merger ULIG FF\,1708+4630 with an unusually high S\'{e}rsic parameter at N=6.  Another bulge dominated object is the IRG FF\,0835+3142.  Otherwise, the majority of the objects seem to be disk-dominated rather than bulge dominated, though the S\'{e}rsic parameters range from 0.7 to 2.3.  

The statistical breakdown for profiles produced in \cite{vei02} for single nucleus objects are that 2\% are best fit by a pure exponential disk, 35\% are best fit by a pure elliptical, and 38\% are fit equally well by both.  There are 10 single-nucleus LIGs in our sample.  Four (40\%) are fit fit well by near exponential shapes, and four (40\%) fall somewhere in between exponential and de Vaucouleurs.  The ``in between" S\'{e}rsic indices indicate that they can be fit by either a pure exponential or a de Vaucouleurs profile but neither is ideal.  The remaining two objects (20\%) have indices that are unusual high, indicating that neither an exponential nor a de Vaucouleurs profile would be a good fit.   We do not find any preference toward elliptical profiles for the LIGs, and instead find more disks, but this may be due to small sample statistics.  We obtain a mean $r_{\frac{1}{2}}$ of 2.9 kpc for the single-nucleus LIGs in $H$ (except for FF1122+4315 which was imaged in $K'$); this is somewhat smaller than the mean half-light radius of 3.5 kpc for ULIGs found by \citet{vei02}.  
Of the IRGs only two are fit well by models. FF1113+5524 has a classic disk+bulge profile and FF\,1712+3205 has a nearly a lenticular shape.  

Boxiness refers to the shape of the isophotes.  Objects with a positive value (``boxy" objects) have 
slightly more square isophotes while those that have a negative value (``disky" objects) are rounded.  
Boxiness tends to be a sign that a galaxy has undergone a recent 
tidal interaction.  Most elliptical galaxies are disky;
elliptical galaxies that are boxy tend to have higher mass to light ratios   
 \citep{kor89}.   Boxiness does not seem to be correlated with infrared luminosity in this sample.

Galaxies that are interacting make up 30\% of the sample, including one ULIG 
and five LIGs.   If we include all objects that are possibly interacting, 
they would make up 80\% of the sample.  Two of the galaxies that do not show 
signs of interaction are IRGs and the other two are LIGs.  So, in agreement 
with previous studies, we observe a trend of toward a higher interaction rate 
at higher luminosities.

FF\,0834+4831, FF\,0835+3142, FF\,1122+4315, and FF\,1519+3520 are identified 
as AGN.  FF\,1519+3520 is an early merger of two galaxies, and possibly a 
dust obscured AGN. FF\,0834+4831 was not imaged.  The other two objects, as 
well as FF\,0819+2707 and FF\,1429+3146, have point-like cores in their 
residuals that are likely to be their active nuclei.  
Except for FF\,0934+4706 which has a PSF like core, and FF\,1318+3250 for 
which we only have an image of the central region, those galaxies below or 
slightly above the demarcation line for starburst galaxies in Fig.~\ref{agnid}
can be modeled with near exponential profiles, and none of them show overt 
signs of tidal interaction.  Conversely, all of the objects that are currently
involved in a tidal interaction are found either in the Seyfert region or in
an intermediate region between the starburst and Seyfert regions.   It is possible
then that every object in the sample that is undergoing a merger has some 
level of nuclear activity.

\section{Notes on Individual Objects}

\subsection{The ULIGS}

\begin{figure}[htb]
\epsscale{0.7}
\plotone{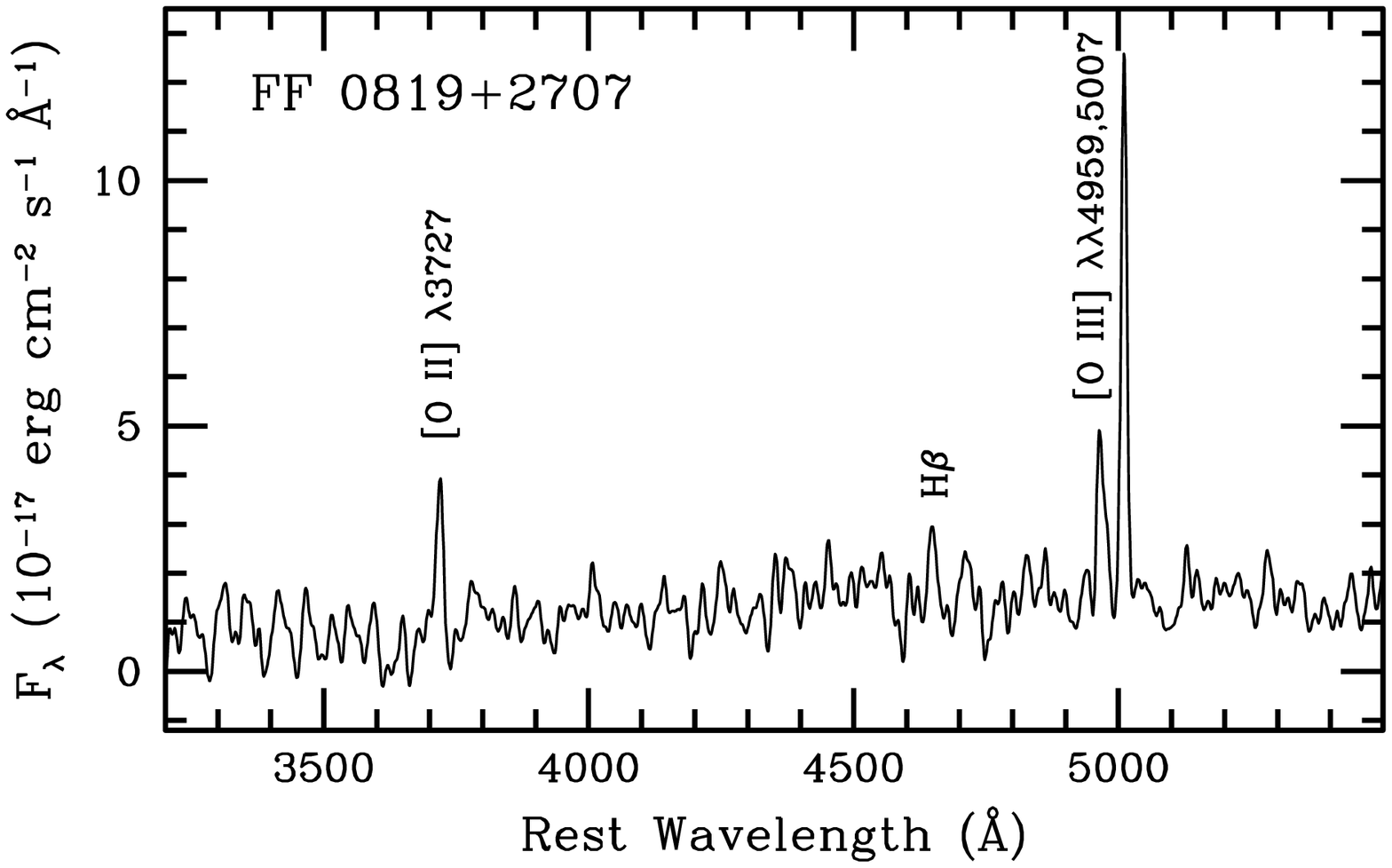}
\caption {Optical spectrum of the $z=0.2613$ ULIG FF\,0819+2707 in rest frame.
\label{spectra}}
\end{figure}

\textbf{FF\,0819+2707} This ULIG is the most powerful object imaged in the sample, with a FIR luminosity of log$(\rm L_{FIR}/L_{\odot}) = 12.5 $.  
Unfortunately, we do not have a flux value for [N\,II] or H$\alpha$ for this object since these lines were redshifted out of the spectra we obtained.  However, the value of log [O\,III]/H$\beta$ = 1.0 for this object indicates it would almost certainly be in the AGN Seyfert range of the BPT diagram.  The spectrum, shown in Fig.~\ref{spectra}, shows a young stellar component.   
No signs of an ongoing merger are evident in the residual, nor is a close companion apparent.  However, the model is not a good fit since it 
undersubtracts at the edges, possibly indicating that the galaxy is perturbed.
The redshift for this object was published by \cite{bra03} after we carried out our observations.  Brand et al. indicate that the object lies within a superstructure of radio galaxies.  

\begin{figure}[htb]
\epsscale{0.7}
\plotone{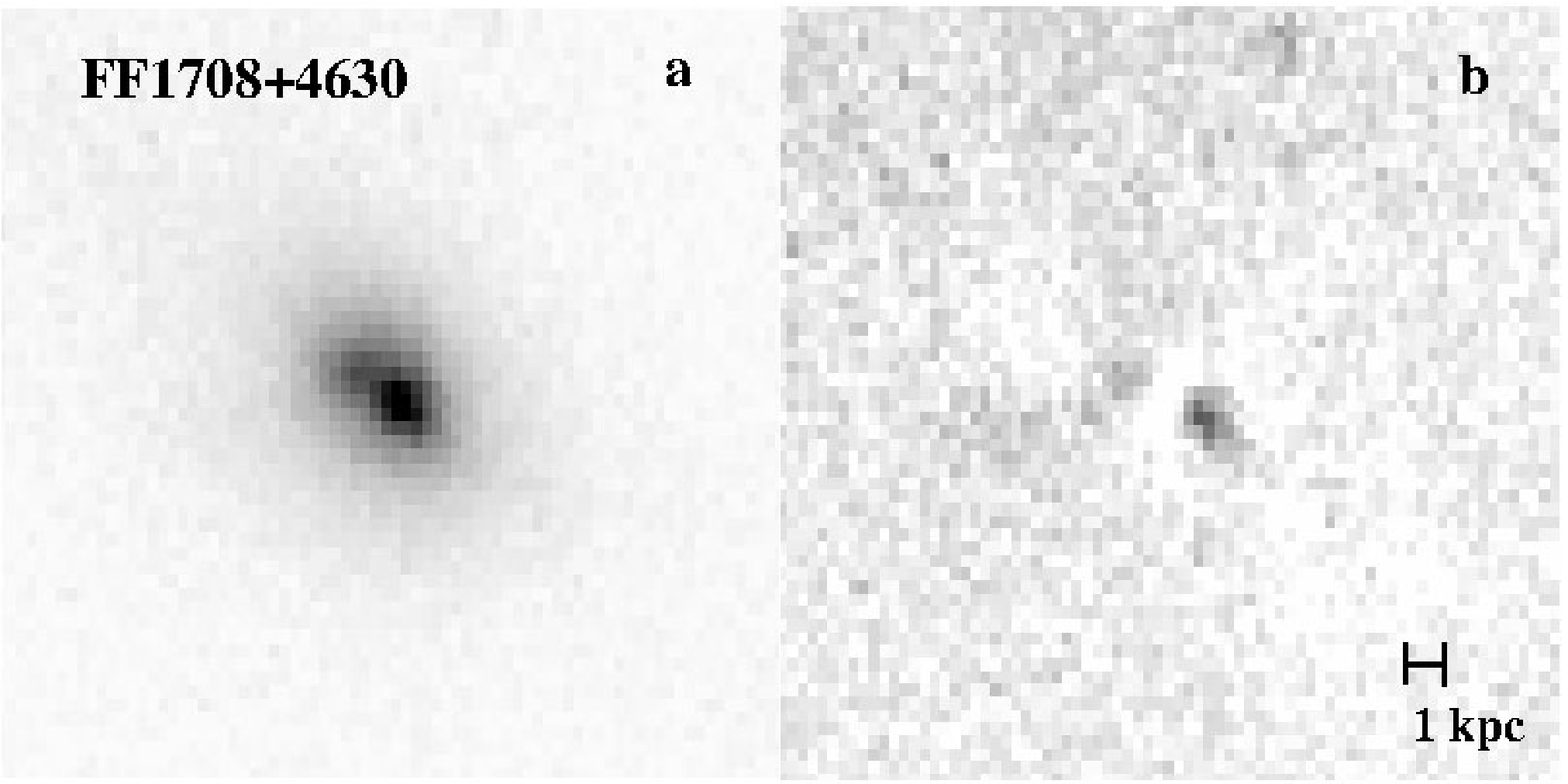}
\caption {Panel $a$ is an image of the $z=0.2630$ ULIG FF\,1708+4630 before subtraction.  Panel $b$ is an image of the object after subtraction, showing a secondary nucleus or merging companion.
\label{1708}}
\end{figure}

\textbf{FF\,1708+4630}  For this galaxy, a merger is evident in Fig.~\ref{1708} after subtraction of a boxy elliptical host with S\'{e}rsic 
parameter N=6.   The merger components are difficult to model.  One of them is probably the core of the elliptical galaxy that is less than 1 kpc in diameter (and correspondingly less luminous).  The intruder seems to be a smaller galaxy about 3 kpc in diameter. It lies only 1.5 kpc (in projection) from the core of the primary galaxy.  The object has a starburst spectrum with ongoing star formation as indicated by strong \ion{O}{2} emission, possibly triggered by the merger. Because of the higher redshift for this galaxy, [N\,II] and H$\alpha$ are also redshifted out of our spectral range.  However, the value of log [O\,III]/H$\beta$ = 0.62 indicates that this object is likely in the AGN region.   The Mg\,I$b$ and \ion{Ca}{2} absorption features are present indicating that an older population is also present.

\subsection{The LIGS}

\textbf{FF\,0825+5216} 
The dominant power source of this log$(\rm L_{FIR}/L_{\odot}) \leq 12.04 $ galaxy is uncertain.  Its spectrum is that of an old population with very weak [O\,II] emission.
The galaxy is less than 10 kpc in diameter in projection and is best modeled by a S\'{e}rsic model with N=2.29.  No signs of a companion are observed.
The model is fairly accurate and there are almost no residuals that are detectable.
The galaxy may have gone through a ULIG phase in the past and is now relaxing to an elliptical shape.  

\textbf{FF\,0839+3626} After subtraction, this moderately luminous LIG reveals a barred spiral (SBc).  The disk is modeled by a S\'{e}rsic profile with N=1.77.   Based on the spectrum, the disk is probably composed of a population of older stars. 
The larger arm is about 2.1 kpc wide and 8.4 kpc long.  This arm is actually visible before subtraction.  The other arm is short and faint.  The bar is about 1.4 kpc wide.  No companion is visible.  The spectrum reveals low ionization emission lines with a strong H$\alpha$ line.   Unfortunately, [O\,III] fell precisely in a small gap between the blue and red detectors so that we cannot obtain an [O\,III]/H$\beta$ ratio.  However, the value of [N\,II]/H$\alpha=-0.16$ indicates that the object may fall in the AGN or transition region.

\textbf{FF\,0841+3557} This object lies on the boundary between LIGs and IRGs.  
The galaxy does not seem to be interacting.  The spectrum indicates it has an old population.  
There are practically no residuals in the subtracted image.  It is best modeled by an exponential profile.  From the morphology it appears to be an ordinary disk galaxy; no bulge is detectable, but this could be due to undersampling. 

\textbf{FF\,0934+4706} The spectrum of this object shows a blue continuum with low ionization emission lines.  Based on its line ratios, the object has a starburst.  This galaxy has a cuspy profile reflecting a high concentration of mass in the center.   It is difficult to model accurately.  It also has a bar-like structure visible in the residual but spiral arms are not apparent.

\textbf{FF\,1110+3130}  This lower luminosity LIG shows a starburst spectrum with a blue continuum. 
The galaxy is fit accurately by a S\'{e}rsic profile with N=0.67, and there are no identifiable residuals.  The shape of this model is near Gaussian (the S\'{e}rsic profile assumes the Gaussian shape when N=0.5).   The galaxy is probably a perturbed disk,  but no companion galaxy is visible. 

\begin{figure}[htb]
\epsscale{0.7}
\plotone{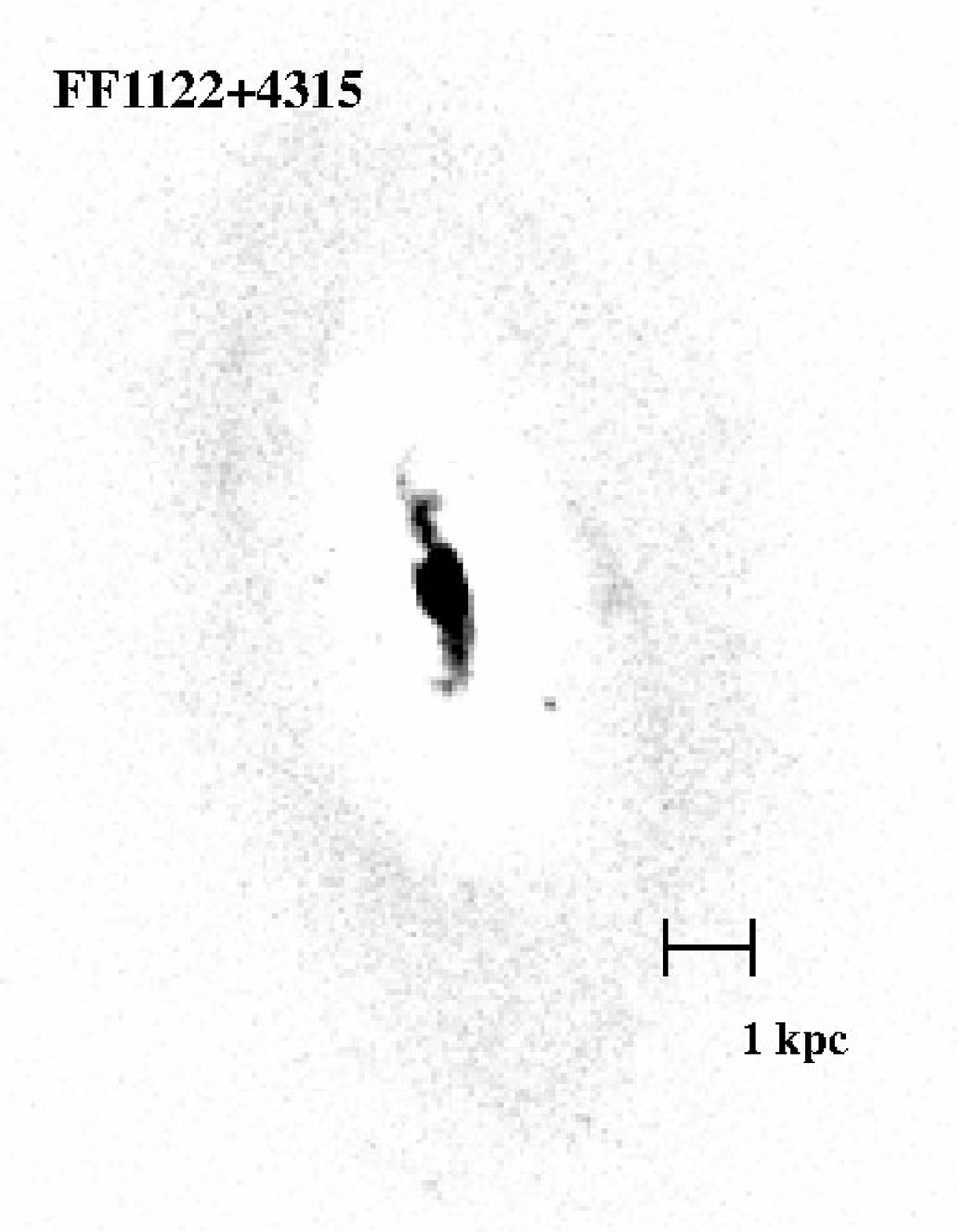}
\caption {Residual image (model subtracted) of the $z=0.1484$ LIG FF\,1122+4315 showing much structure near the AGN nucleus. 
 \label{ff1122}}
\end{figure}

\textbf{FF\,1122+4315} This luminous LIG appears to be a spiral galaxy with a bulge-like component that extends over most of the image.  The spectrum indicates it is an AGN Seyfert galaxy.  An elliptical shape with a S\'{e}rsic index of N=3.09 was used to model the bulge component. The residual image 
(Fig.~\ref{ff1122}) shows much structure including a possible warped disk 
and/or tidal debris, although some effects of over-subtraction are evident.  
A small knot to the southwest could be the core of an interacting dwarf 
galaxy or a bright knot of star formation.

\textbf{FF\,1316+2511} This is a strongly perturbed LIG.  The S\'{e}rsic exponent, while unusually high (N=10), seems to produce an accurate model of this cuspy galaxy; practically no residuals are seen after subtraction.   The galaxy is spherical in shape (E0).  It is boxy, but no other signs of recent interaction are apparent.  The spectrum indicates that an old population of stars is present.

\begin{figure}[htb]
\epsscale{0.7}
\plotone{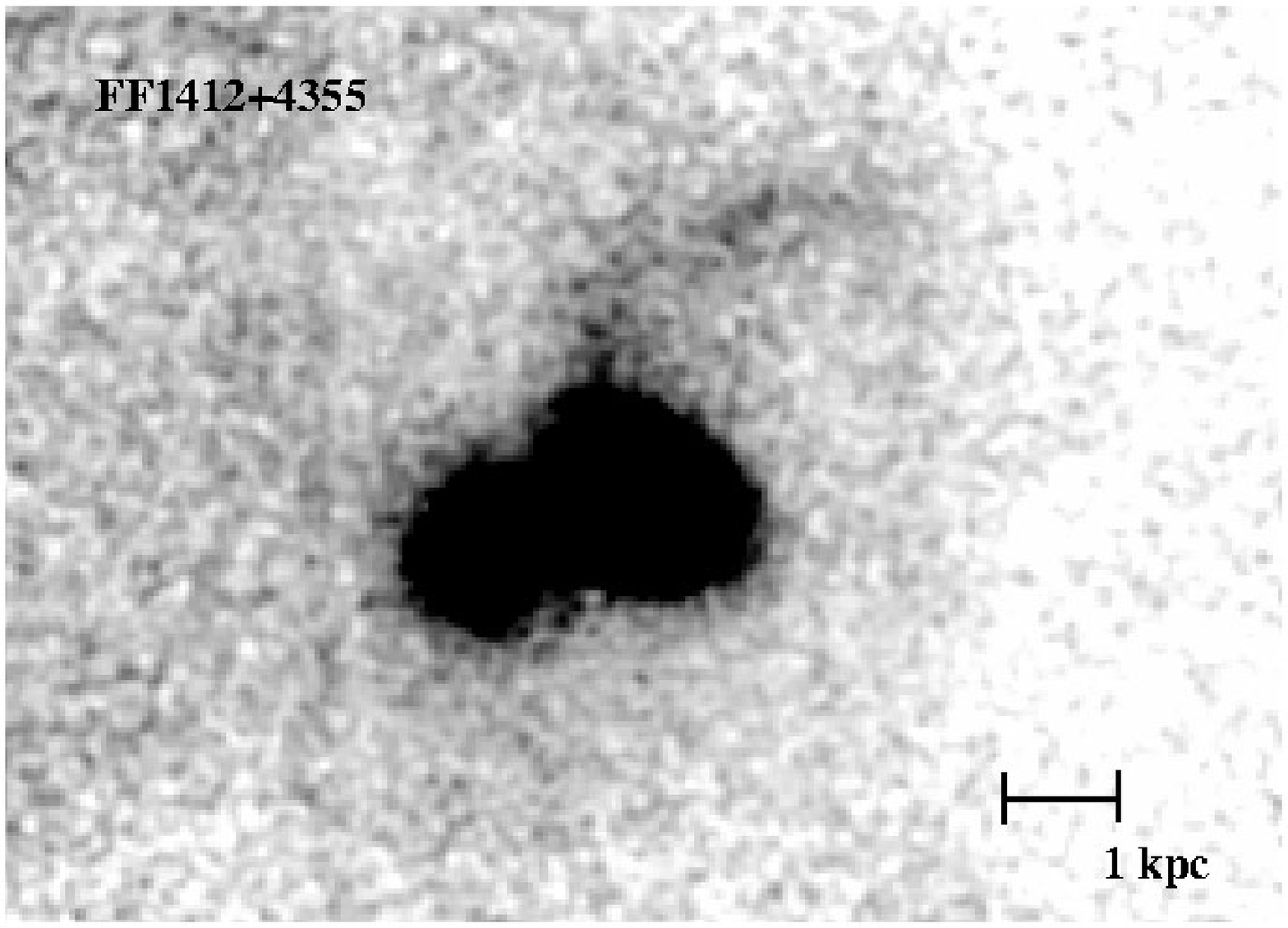}
\caption {Two merging disk galaxies form the $z=0.1353$ LIG FF\,1412+4355.  The image has been smoothed using a gaussian with $\sigma=1$ pixel to highlight the tidal tail and debris.
\label{ff1412}}
\end{figure}

\textbf{FF\,1412+4355} This object is a likely starburst galaxy though it lies slightly above the demarcation line in Fig.~\ref{agnid}.  Figure \ref{ff1412} shows that the galaxy is an early merger of two disk galaxies with tidal tails and extended debris.  The larger spiral galaxy is well modeled by an exponential profile, and it has a very long tidal tail.   The galaxies are $\sim$1.8 kpc apart in projection. The smaller companion is modeled well by a boxy S\'{e}rsic with N=1.7, and shows practically no residuals after subtraction, while the larger galaxy reveals a distinct core 0.8 kpc in diameter and a diffuse cloud of debris.

\begin{figure}[htb]
\epsscale{0.7}
\plotone{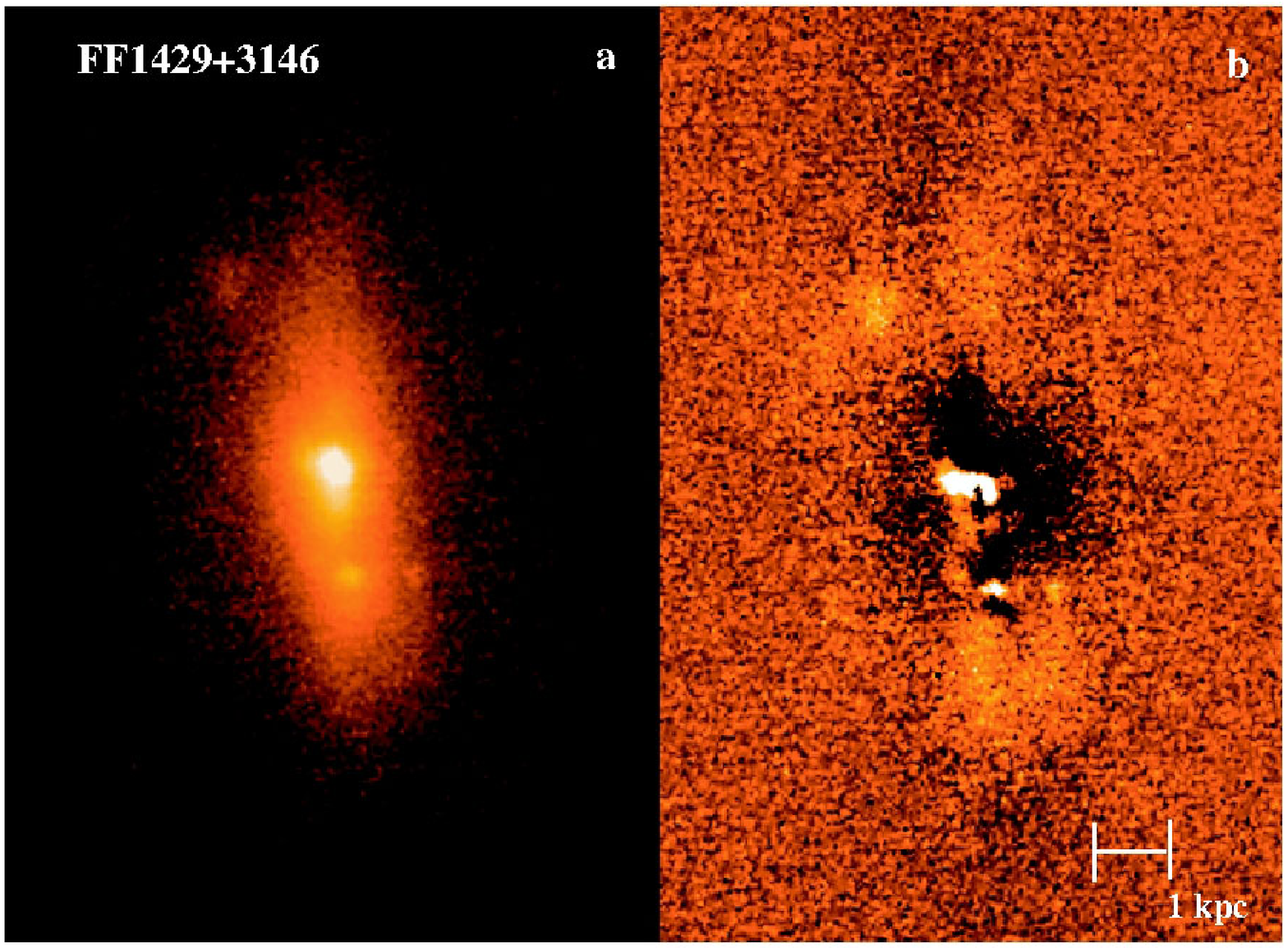}
\caption {Keck NIRC-2 Narrow Camera image of the $z=0.1806$ LIG FF\,1429+3146, displayed in a log scale.  Panel $a$ shows the galaxy before subtraction and panel $b$ after subtraction.  The dark regions are an artifact of over-subtraction due to the highly perturbed morphology and multiple components of this galaxy.
\label{ff1429}}
\end{figure}

\textbf{FF\,1429+3146}  The emission line ratios for this object indicate it is somewhere between the AGN and starburst regions on the BPT diagram.  However given the morphology of this galaxy, and a possible faint broad emission line component in its optical spectrum, it could be an obscured AGN. The galaxy is clearly undergoing an early merger and contains a young population of stars.  
We obtained two separate images of this galaxy using the narrow and wide 
NIRC-2 cameras. Figure \ref{ff1429} shows the narrow camera image.  Each image was 
modeled independently; the model parameters determined for each of these images are generally consistent with each other, although the (higher resolution) narrow camera image yields a better fit.  The parameters listed in Table~\ref{parameters} correspond to the narrow camera image. 
FF\,1429+3146 has four distinct features.  The two major components are a large cuspy S\'{e}rsic profile and a smaller merging companion to the south also with a S\'{e}rsic profile.  There is a dwarf companion or star forming region to the northeast and possibly one to the west.  All of these components are visible before subtraction.   The north galaxy has a S\'{e}rsic parameter of N$\sim$2, while the south galaxy has a S\'{e}rsic parameter of N$\sim$1.1 corresponding to an exponential shape.  
The models for both major components show diskiness.   In the residual images, a core about 0.5 kpc across corresponding to the larger galaxy can be seen.  This is possibly an active nucleus.     The galaxies are about 1.5 kpc apart in projection.  The unusual shape of this galaxy could indicate it is undergoing a major merger.  In addition there may also be a minor merger with a smaller companion making it a multiple merger.  There seems to be some dust present to the north and south.

\textbf{FF\,1517+2800}  This galaxy has a bright S\'{e}rsic profile with N=2.17.  The galaxy also has a faint S\'{e}rsic component $\sim$0.5 kpc off center with N=0.31.  This unusual shape is similar to a Gaussian, but steeper. The subtracted image shows a core $\sim$1~kpc in diameter and some other residuals that may be a tidal tail. Some debris to the south of the obvious core may be another core that is part of a former companion. The spectrum of this galaxy is heavily reddened and shows strong emission lines.  Emission line ratios indicate that the object is probably a starburst galaxy, although it lies in the region between starbursts and AGNs in Fig.~\ref{agnid}.

\textbf{FF\,1519+3520} This system is best modeled by disk components.  The larger component has a S\'{e}rsic index of N=1.44 while a smaller component on the southeast has an N=0.91 index and is $\sim$4 magnitudes less luminous.  There is a bright, resolved core about 1.5 kpc in diameter roughly at the position of the smaller galaxy.  It is presumably merging with the primary source.   There are features that could be tidal tails to the north and south.  The spectrum indicates that the stellar population is older.   Its FIR luminosity puts it at the lower end of the LIG scale. 

\textbf{FF\,1709+5220} This moderately luminous LIG is likely to be powered by a starburst.  The galaxy is best modeled by a S\'{e}rsic profile of index N=0.66; this index could be low due to undersampling.  The image is not deep enough to determine if there is a bulge component.  The galaxy may have a companion of similar radius and magnitude, with a S\'{e}rsic index of N=0.48 that lies 16\arcsec\ away in the image.  Since we do not have a redshift for this object, we cannot determine whether it is a projected galaxy or a true companion.

\textbf{FF\,1725+4559} This low luminosity LIG is best modeled by an exponential profile.  It is slightly boxy but otherwise appears normal.  The subtracted image reveals a possible core about 1~kpc long and 0.5~kpc wide.  It lies at the same position angle as the disk and it could be a bulge that is too small to model.  In that case this galaxy would seem to be a normal disk + bulge galaxy.

\subsection{The IRGs}
\textbf{FF\,0835+3142}  This object contains an older population of stars.  It can be fit by a boxy elliptical host but is difficult to model accurately.  The subtracted image reveals a possible unresolved core.

\textbf{FF\,1113+5524}  This object seems to be a normal spiral galaxy.    Its spectrum shows some some star forming activity. The residuals show that it is a prime example of an Sbc spiral galaxy.  A prominent bar and two arms nearly 1~kpc wide each can be seen.  

\textbf{FF\,1138+4405}  This object can be fit with a near exponential profile. The residual image reveals a very unusual oblong shaped core.  The spectrum of this object is reddened showing very strong emission (the strongest of the sample) that classifies this object as a starburst, and an underlying old stellar population.  

\textbf{FF\,1318+3250}  The image of this galaxy in Fig.~\ref{msgals} shows only the central bar of the galaxy.  The galaxy has two faint spiral arms visible only in optical images.  The bulk of the NIR flux comes from the bar, which is $\sim$15~kpc long in projection.  The bulge is $\sim$4~kpc in diameter.  The position angles of the bar and bulge are offset by 20 degrees.  The spectrum of this galaxy shows some star formation based on the presence of [O\,II] emission, but the bulk of the stellar population is probably old.  No AGN is apparent.  

\textbf{FF\,1712+3205}  This galaxy can be modeled by an N=2.5 disky S\'{e}rsic, which is roughly a lenticular shape. The spectrum shows weak H$\alpha$ and [O\,II] emission.

\section{Summary\label{summary}}

The results from this study indicate that the technique of cross-correlating 1.4 GHz and far-infrared fluxes selects many perturbed galaxies, with frequent early mergers and merger remnants.   It selects a high number of starburst galaxies at redshifts $0.1<z<0.3$.  A few type-2 AGN are picked up using this technique but they are much less common.

While our sample is small and we are dealing with small number statistics, our
results confirm several trends observed before: 1) ULIGs are almost invariably
mergers or interacting galaxies; 2) the fraction of LIGs undergoing mergers is
significantly less; 3) objects with higher FIR luminosity are more likely to
contain AGN.  

We find, on the other hand, a larger fraction of exponential or
near-exponential profiles (nearly half of the sample) than in previous surveys.
As discussed in Section 4.1, many of these surveys have based their 
classification of profiles on one dimensional fits which can be subject to
errors due to isophote twists or other small perturbations.   It is possible
that the true fraction of objects with exponential profiles may be larger,
as suggested by our study.  However, because of the small size of our sample
this is simply speculation at this stage.


We do confirm, however, that modeling in two dimensions is very effective in
highlighting features that could be easily missed by visual inspection or
one dimensional fitting.  Residual images reveal details 
such as double nuclei, dust lanes, tidal debris, and secondary cores that
allow us to identify more accurately those objects that are perturbed.


Our results show the effectiveness of using adaptive optics systems in
combination with two dimensional modeling to 
study morphologies of infrared galaxies.
Future morphological surveys of this kind can be done with success for other 
types of interesting galaxies.

\acknowledgments
We gratefully acknowledge Chien Peng for helpful discussions about galaxy 
fitting.  We thank both of the anonymous referees who reviewed our paper 
for their helpful
comments and suggestions.  Some of the data presented herein were 
obtained at the W.M. Keck Observatory, which is operated as a scientific 
partnership among the California Institute of Technology, the University 
of California and the National Aeronautics and Space Administration. The 
Observatory was made possible by the generous financial support of the
W.M. Keck Foundation. 
This work was supported in part under the auspices 
of the U.S.\ Department of Energy, National Nuclear Security 
Administration by the University of California, Lawrence Livermore 
National Laboratory under contract No. W-7405-Eng-48.

\end{document}

%% file: tab1.tex
\begin{deluxetable}{lcccccccc}
\tabletypesize{\small}
\tablecaption{The Sample\label{sample}\tablenotemark{a}}
\tablehead{\colhead{Object} & \colhead{$\alpha$} & \colhead{$\delta$} &  \colhead{$z$} & \colhead{$f_{60\mu m}$} & \colhead{$f_{100\mu m}$\tablenotemark{b}} & \colhead{DL}  & \colhead{$f_{1.4GHz}$} & 
\colhead{log $L_{FIR}$\tablenotemark{c}} \\
 \colhead{Name} 
& \colhead{(J2000)} & \colhead{(J2000)} & \colhead{} & \colhead{(Jy)} & \colhead{(Jy)} & \colhead{(Mpc)}& \colhead{(mJy)} & \colhead{($L_{\odot}$)}  \\
 \colhead{(1)} 
& \colhead{(2)} & \colhead{(3)} & \colhead{(4)} & \colhead{(5)} & \colhead{(6)} & \colhead{(7)}& \colhead{(8)} & \colhead{(9)} }
\startdata	
FF0819+2707 	&   08 19 16.8&     27 07 34      	&0.2613	&0.74		&1.07	&1314		&5.97				        &12.50  \\	
FF0825+5216	 &08 25 34.5&     52 16 42 	&0.1726	&0.73		&(0.65)	&823.6		&2.4				        &(12.03)  \\	
FF0835+3142	&	08 35 51.6&     31 42 00 	&0.0483	&0.36		&(0.30)	&211.7		&1.02				        &(10.53)  \\	
FF0839+3626	&08 39 50.5&     36 26 57 		&0.0961	&0.48		&1.14	&435.9		&3.12				        &11.44  \\	
FF0841+3557	&08 41 30.9&     35 57 45 		&0.0502	&0.57		&1.68	&220.3		&2.72				        &10.98  \\	
FF0934+4706	&09 34 04.0&     47 06 02 		&0.1207	&0.33		&(0.58)	&556.8		&1.46				        &(11.43)  \\	
FF1110+3130	&11 10 02.2&     31 30 02 		&0.1171	&0.29		&(0.20)	&538.9		&1.06				        &(11.23)  \\
FF1113+5524	&11 13 38.6&     55 24 41 		&0.0382   &0.71		&1.50	&166.1		&3.33				        &10.75  \\	
FF1122+4315 	&11 22 03.6&     43 15 56		&0.1463	&0.46		&0.93	&686.4		&3.47				        &11.79  \\	
FF1138+4405	&11 38 35.5&     44 05 28 		&0.0359	&0.68		&0.86	&155.9		&3.35				        &10.59  \\	
FF1316+2511	&13 16 42.1&     25 11 56 		&0.1459	&0.34		&(0.59)	&684.4		&1.87				        &(11.63)  \\	
FF1318+3250	&13 18 24.3&     32 50 41 		&0.0367	&0.51		&0.84	&159.4		&2.68				        &10.53  \\	
FF1412+4355	&14 12 29.6&     43 55 55 		&0.1332	&0.59		&0.85	&619.6		&1.68				        &11.75  \\	
FF1429+3146 	&14 29 56.5&     31 46 02		&0.1761	&0.18		&(0.25)	&842.1		&1.09				        &(11.50)  \\ 
FF1517+2800	&15 17 52.8&     28 00 50 		&0.1016	&0.39		&(0.60)	&462.6		&2.24				        &(11.32)  \\	
FF1519+3520	&15 19 58.4&     35 20 37 		&0.1098	&0.24		&(0.49)	&502.8		&2.1				        &(11.23)  \\	
FF1708+4630	&17 08 54.0&     46 30 46 		&0.2630	&0.29		&0.90	&1323		&2.99				        &12.26  \\
FF1709+5220  &  17 09 00.8   &    52 20 03       &0.1689	&0.20                 &(0.52)	&804.0  	&1.18	        		        &(11.61)\\
FF1712+3205	&17 12 07.9&     32 05 33 		&0.0372	&0.27		&(0.66)	&161.7		&1.19				        &(10.34)  \\	
FF1725+4559	&17 25 00.3&     45 59 43 		&0.0625	&1.10		&1.03	&276.8		&4.43				        &11.26  \\
\hline FF0834+4831  &08 34 46.8&  48 31 39    & 0.1735 &0.30                 & (0.80) & 828.3  &2.13  &  (11.84) \\
FF1439+3232\tablenotemark{c}  &14 39 16.9& 32 32 39  &0.2502  &  \nodata  & \nodata  &  1250  & 1.21  &  \nodata \\	
FF1601+4514	 & 16 01 56.6& 45 14 03            &0.0969	&0.58		&1.12	&439.8		&3.35			    	&11.50\\	
FF1621+2214	 & 16 21 08.1 &  22 14 08          &0.0843	&0.29		&0.60	&379.2		&1.37			    	&11.08\\	
FF1651+3001	& 16 51 22.6    &     30 01 04     &0.0592	&0.55		&0.96	&275.4		&4.32			    	&11.02\\
FF1656+2644	&  16 56 46.5&   26 44 57       &0.1193	&0.44		&(0.69)	&574.8		&1.16			    	&(11.55)\\
FF1721+2951	& 17 21 43.7    &     29 50 59     &0.1052	&0.23		&(0.48)	&480.2		&1.07			    	&(11.19)\\ 
FF1723+3845	& 17 23 29.7    &   38 45 12       &0.0377	&0.22		&(0.28)	&163.9		&1.04			   	&(10.15)\\	
\enddata 
\tablenotetext{a}{Horizontal line divides imaged and non-imaged samples.}
\tablenotetext{b}{Parentheses indicate upper limits.}
\tablenotetext{c}{$IRAS$ data for this object suffers contamination from a nearby M6 star.}
\end{deluxetable}

%% file: tab2.tex
\begin{deluxetable}{lccccccc}
\tabletypesize{\small}
\rotate
\tablewidth{575pt}
\tablecaption{Journal of Observations\label{journal}}
\tablehead{\colhead{Object} 
 & \colhead{Scale (kpc/$\arcsec$)} & \colhead{GS $V$} & \colhead{Separation (\arcsec )} & \colhead{PA (deg)} &\colhead{Exp. Time (s)} &\colhead{FWHM (\arcsec ) }  &
 \colhead{Date Obs.} \\
\colhead{(1)} 
& \colhead{(2)} & \colhead{(3)} & \colhead{(4)} & \colhead{(5)} & \colhead{(6)} & \colhead{(7)} & \colhead{(8)}  }  
\startdata	
FF0819+2707 & 	4.028&	   11.5&	17.9&\phn64.8 &	12$\times$300	&	  0.49  &	25 Jan 2003\\
FF0825+5216 & 	2.904&     11.9&	30.2& \phn61.2 &	14$\times$300	& 0.49  &	25 Jan 2003\\
FF0835+3142 & 	0.981&     12.5&	37.9&  286.1 &	10$\times$300	&	  0.49  &	25 Jan 2003\\
FF0839+3626 & 	1.809&     11.7 &	31.3&  250.6 &	\phn6$\times$300	& 0.15  &	03 Apr 2002\\
FF0841+3557 & 	0.968&     12.5 &	31.6&  151.6 &	\phn3$\times$300	& 0.49  &	25 Jan 2003\\
FF0934+4706 & 	2.149&     12.8&	34.8&\phn\phn5.4 &12$\times$300       &0.31     &	03 Apr 2002\\
FF1110+3130 & 	2.134&     12.6&	43.6&  127.2          &	\phn6$\times$600& 0.80  &	19 Mar 2003\\
FF1113+5524 & 	0.762&  \phn7.5&	37.0&  295.9           & \phn7$\times$300  & 0.49 & 25 Jan 2003\\
FF1122+4315\tablenotemark{a} &2.563&12.7& 27.3&267.9            & \phn25$\times$60 &     0.068 &25 May 2002\\
FF1138+4405 & 	0.732&  \phn9.7&	29.6&  258.9              &	10$\times$300	& 0.43	&03 Apr 2002\\
FF1316+2511 & 	2.527&     11.2&	40.6&\phn79.9           &	\phn8$\times$300&0.60	&03 Apr 2002\\
FF1318+3250 & 	0.719&     10.5&	34.0&  115.5          &    \phn6$\times$300 &0.49&       03 Apr 2002\\
FF1412+4355 & 	2.371&  \phn9.8&	27.4&  350.7          &	10$\times$300	&	0.39	& 03 Apr 2002\\
FF1429+3146\tablenotemark{a} & 3.012&11.9&30.6&\phn\phn8.2        &	10$\times$120 & 0.092  	&25 May 2002\\
	&	&	&	&                                   &\phn5$\times$120 & 0.068 & 25 May 2002 \\
FF1517+2800 & 	1.848&     11.9&	30.0& \phn13.6        &	\phn6$\times$300 &  0.42&	03 Apr 2002\\
FF1519+3520 & 	1.979&     11.1&	36.2& 188.3       &	\phn6$\times$300 & 0.29	&	19 Mar 2003\\
FF1708+4630 & 	4.024& \phn9.1 &	27.2& 161.2     &	\phn7$\times$300 & 0.44	&	03 Apr 2002\\
FF1709+5220 & 	2.853&     11.8 &	27.2& 237.0         &	11$\times$300	& 0.14	&	04 Sep 2004\\
FF1712+3205 & 	0.729&     11.1&	43.0& 254.1         &	\phn5$\times$300 & 0.17 &	14 Aug 2003\\
FF1725+4559 & 	1.189&     11.9&	27.4& 187.5         &	16$\times$300	& 0.63	&	05 Sep 2003\\
 \enddata 
\tablenotetext{a}{Galaxies observed in $K'$ with the Keck II telescope}
\end{deluxetable}

%% file: tab3.tex
\begin{deluxetable}{lccccrrccc}
\tabletypesize{\small}
\tablewidth{575pt}
\tablecaption{Model Parameters \& Morphologies \label{parameters}}
\tablehead{\colhead{Object} 
 & \colhead{MAG} & \colhead{N} & \colhead{$r_{\frac{1}{2}}$ (kpc)} & \colhead{$\varepsilon$} & \colhead{PA (deg)} & \colhead{Boxy/Disky} 
 & \colhead{Fit Type} & \colhead{Nucleus} & \colhead{Interact?}
\\ \colhead{(1)} 
& \colhead{(2)} & \colhead{(3)} & \colhead{(4)} & \colhead{(5)} & \colhead{(6)}& \colhead{(7)}& \colhead{(8)}& \colhead{(9)}& \colhead{(10)}}
\startdata
FF0819+2707 	&		25.1	&	2.7	&	5.5	&	0.52	&	76.0	&	$\sim0$	&	S\'{e}r  	&     single 	&    ?        \\
FF0825+5216 	&		25.0	&	2.3	&	0.90	&	0.44	&	$-$64.2	&	0.27		&	 S\'{e}r &  single   	&     ?     \\
FF0835+3142 	&		23.3	&	4.0	&	6.8	&	0.80	&	$-$61.0	&	0.68	         &	 S\'{e}r &	single  	&     ?     \\
FF0839+3626 	&		20.0	&	1.8	&	3.1	&	0.52	&	$-$34.4	&	$-$0.31	&	 S\'{e}r &	single     	&     N   \\
FF0841+3557 	&		23.6	&	1.0	&	4.6	&	0.26	&	9.0		&	$-$0.83	&	Exp  	&	single     	&      N    \\
FF0934+4706 	&		19.0	&	7.2	&	3.0	&	0.74	&	82.8		&	$-$0.54	&	 S\'{e}r& 	single     	&     ?   \\
FF1110+3130 	&		22.4	&	0.7	&	2.9	&	0.78	&	23.3		&	$-$0.15	&	 S\'{e}r& 	single     	&     ?  \\
FF1113+5524 	&		21.8	&	5.0	&	2.3	&	0.47	&	17.3		&	0.15		&	 S\'{e}r& 	single     	&     N \\
		&		30.5	&	1.0	&	8.7	&	0.64	&	8.0	 	&	$-$0.81	&	Exp  	&	               	&       \\
FF1122+4315 	&	    \phn8.8	&	3.1	&	6.9	&	0.53	&	13.1		&	$-$0.16	&	S\'{e}r &	single    	&     Y  \\
FF1138+4405 	&		20.5	&	1.16	&	1.7	&	0.84	&	34.6	&	0.02	&	  S\'{e} r & 	single     	&     Y \\
FF1316+2511  &		20.6	&	10.0	&	2.8	&	0.84	&	$-$33.3	&	0.36		&	 S\'{e}r &	single      	&    ? \\
FF1318+3250\tablenotemark{a} 	& \nodata & \nodata	&	\nodata	& \nodata		& \nodata		&	\nodata		&	Bar & single & N      \\
FF1412+4355 	&		23.0	&	1.7	&	0.67	&	0.90	&	$-$43.0	&	0.47	&	  S\'{e}r & 	double      &     Y \\
			&	21.6	&	1.0	&	1.9	&	0.09	&	85.9		&	$\sim0$	&	Exp   &         &      \\
FF1429+3146 	&		15.2	&	2.3	&	5.9	&	0.52	&	$-$1.4	&	$-$0.51	&	   S\'{e}r&	mult        &    Y  \\
			&	19.1	&	1.1	&	1.2	&	0.68	&	24.1		&	$-$0.33	&	   S\'{e}r&	                   &     \\
FF1517+2800 	&		20.7	&	2.2	&	0.90	&	0.61	&	$-$22.6	&	$-$0.23	&	  S\'{e}r &	single        &  Y    \\
			&	22.0	&	0.3	&	2.8	&	0.49	&	44.7		&	0.08		&	  S\'{e}r &	                   &      \\
FF1519+3520 	&		20.3	&	1.4	&	2.8	&	0.68	&	$-$35.4	&	$-$0.20	&	  S\'{e}r &	double      &  Y    \\
			&	23.7	&	0.9	&	0.72	&	0.85	&	86.7		&	$-$0.33	&	  S\'{e}r &	                   &      \\
FF1708+4630	&		21.3	&	6.0	&	4.0	&	0.26	&	62.8		&	0.31		&	  S\'{e}r &	double       &   Y     \\
FF1709+5220     &		23.1	&	0.7 	&	2.3	&	0.34 &	35.7		&	$-$0.25	&	  S\'{e}r &	single         &  ?     \\
FF1712+3205	&		20.4	&	2.5	&	1.4	&	0.25	&	$-$79.9	&	$-$0.95	&	  S\'{e}r &	single         &   ?     \\
FF1725+4559	&		20.8	&	1.1	&	1.4	&	0.89	&	30.8		&	0.02		&	   S\'{e}r&	single         &    ?     \\
\enddata 
\tablenotetext{a}{The model parameters for this object are not meaningful since only the central bar was imaged.}
\end{deluxetable}